# Relativistic Achilles

*Dr. Fabrice Leardini*

*Dipartimento di Fisica, Sapienza Università di Roma*

*Piazzale Aldo Moro, 5, 00185, Roma (Italy)*

*e-mail: fabrice.leardini@roma1.infn.it*

**ABSTRACT**

This manuscript presents a problem on special relativity theory (SRT) which embodies an apparent paradox relying on the concept of simultaneity. The problem is represented in the framework of Greek epic poetry and structured in a didactic way. Owing to the characteristic properties of Lorenz transformations, three events which are simultaneous in a given inertial reference system, occur at different times in the other two reference frames. In contrast to the famous twin paradox, in the present case there are three, not two, different inertial observers. This feature provides a better framework to expose some of the main characteristics of SRT, in particular, the concept of velocity and the relativistic rule of addition of velocities.

## I. A MYTHICAL RACE

Now when the child of morning, rosy-fingered Dawn, appeared, a group of men went to the beach, in the coast of Troy. Achilles and his friend, Patroclus, were going to run a race under the judgment of their companions. Once arrived, they laid a string of one stadium in length on the sand and sat down on some rocks close to one of the ends of the string. Firstly, they ordered Patroclus to run along the string and determined the time employed to cover the distance of one stadium. Secondly, Achilles completed the distance marked by the string, employing half of Patroclus' time. Next, they disposed another string, identical to the first one, in series, and stipulated that Achilles should cover the distance of two stadiums, whereas Patroclus should start from the intersection of the two strings, covering only the distance of one stadium. At sunrise, with the first rays of light, the two runners of comely feet started the race simultaneously. The observers on the rocks agreed that the arrival was also simultaneous. However, Patroclus and Achilles knew that one of them reached the rocks first. Both of them were inspired by great Hermes, who shined his bright stick close to Apollo, over the oceanic horizon. The god conferred to them amazing speeds, close to the limit imposed by the laws of space and time. By travelling at these speeds, both runners perceived the victory of Achilles.

## II. UNCOVERING THE APPARENT PARADOX

Since our scholar period we have become used to the description of movement given by classical relativity. This theory is based on the idea of universal scales to measure distances and time intervals, valid for all inertial reference systems (*i.e.*, those reference systems moving at constant relative velocities with respect to each others). Based on this fundamental idea, classical relativity provides a set of rules to transform space and time coordinates of different inertial observers. By using these rules, which are known as Galileo's transformations, we will analyze the race from the point of view of each one of the different observers participating in the scene: the men on the rocks, Patroclus and Achilles. However, we will find that the encounter of Achilles, Patroclus and the men on the rocks is simultaneous for all the observers, contrary to that stated in the myth.

To solve the paradox, we must take into account the figure of Hermes, the god of athletics, who was also associated with the fastest visible planet, Mercury, in ancient Greek mythology. Hermes conferred to Patroclus and Achilles very high speeds, close to the limit of light, relative to the observers sitting on the rocks. Therefore, to analyze the description of the race given by the two epic runners, we shall make use of special relativity, which is currently the most adequate theory for comparing the description of physical reality given by different observers moving at great relative velocities.[1] Within this theory, space and time have no longer the universal validity mentioned above. Instead of that, it is the speed of light ($c$) that is conserved by the rules of transformation of space and time coordinates of different inertial frames. These rules are known as Lorenz transformations. One of the consequences of these transformations is the space contraction experienced by relativistic observers. This implies that the length of the strings will be contracted for Patroclus and even more for Achilles as compared to the observers on the rocks. On the other hand, to calculate the relative velocities of Achilles, Patroclus and the rocks in the corresponding reference systems we must use the relativistic rule of addition of velocities. This relationship is not equal to the simple sum of relative velocities expressed in classical physics. Instead is subject to the limits on the speed of every object in any inertial frame imposed by special relativity and represented by the universal constant *c*. The key to solving the paradox embodied in the myth relies in this fundamental relationship.

**IIA. Analytical description of the race within the framework of classical relativity**

In what follows, we will analyze the race from the point of view of the three reference systems described in the scene and within the framework of classical relativity.

For the observers on the rocks (inertial reference system *S*), Patroclus started the race from the point $x_P(0)=-l$, running with an uniform velocity $v_P=u$, while Achilles did it from the point $x_A(0)=-2l$, running with constant velocity $v_A=2u$ (see Fig.1a). By taking the race start as the origin of time, their respective equations of movement can be written as:

$$x_P(t) = -l + ut \qquad (1.1)$$
$$x_A(t) = -2l + 2ut \qquad (1.2)$$

Eqs. (1.1) and (1.2) show that the distance of Achilles to the finish line was twice that of Patroclus, except at the instant of the arrival, which was simultaneous for both runners, at time $t_f$:

$$t_f = l/u \qquad (1.3)$$

From the point of view of Patroclus (inertial reference system *S'*), the rocks approached him from point $x_R'(0)=l$ at a constant velocity $v_R'=-u$, while Achilles did so from point $x_A'(0)=-l$, at velocity $v_A'=u$ (see Fig.1b). Accordingly, the equations of movement of the rocks and Achilles observed by Patroclus are given by:

$$x_R'(t) = l - ut \qquad (2.1)$$

$$x_A'(t) = -l + ut \qquad (2.2)$$

These equations imply that the distances of the rocks and Achilles to Patroclus were the same at every time and they met at time $t_f$.

In Achilles' reference system (inertial reference system $S''$), Patroclus approached from point $x_P''(0)=l$, at uniform velocity $v_P''=-u$, whereas the rocks did the same from point $x_R''(0)=2l$, at velocity $v_R''=-2u$ (see Fig.1c). As a consequence, the corresponding equations of movement are given by:

$$x_P''(t) = l - ut \qquad (3.1)$$

$$x_R''(t) = 2l - 2ut \qquad (3.2)$$

Again, these equations imply that the rocks, Patroclus and Achilles met simultaneously at time $l/u$.

The graphical representation of the corresponding equations of movement perceived by each one of the different observers is plotted in Fig.2. The relative movements perceived in the three reference systems present a nice symmetry. Besides, it is clearly shown that the encounter of Achilles, Patroclus and the rocks is isochronous and simultaneous for all of them. Within the framework of classical relativity, the myth is misinterpreted.

**IIB. Analytical description of the race within the framework of special relativity**

In what follows, we will analyze how the relative movement would be perceived by each one of the different observers according to the laws of special relativity.

The observers in the inertial system *S* were static relative to the strings. They measured their lengths (*l*) as well as the velocities of Patroclus (*u*) and Achilles (*2u*). To them, Achilles covered twice the distance of Patroclus, and he ran twice fast. As a consequence, the equations of movement of both runners are the same as those written above (Eqs.1.1 and 1.2). In reference frame *S*, Achilles and Patroclus arrived simultaneously to the rocks after time *l/u*. This far the only new condition imposed by special relativity is that the speeds of both runners must be lower than *c*, hence:

$$u < \frac{c}{2} \qquad (4)$$

To analyze the scene from the points of view of Patroclus and Achilles, we will assume that both runners were in rest relative to the rocks at *t=0*. They acquired uniform velocities with respect to the rocks instantaneously due to infinite accelerations of the Dirac delta type. Obviously, these accelerations could only be provided by the divine inspiration of Hermes. Under this hypothesis, we can choose the same time origin in the three inertial frames. Both Patroclus and Achilles perceived that the rocks and each other started approaching simultaneously with the first rays of light. This supposition is analogous to that implicit in the well known *twin paradox*. In that case, we also take the same time origin in the inertial frames of both brothers.

Patroclus observed the rocks advancing from the extreme of the string laid in front of him, whose length (*l'*) contracted with respect to the length measured in *S*, specifically:

$$l' = \sqrt{1 - \frac{u^2}{c^2}}\, l \qquad (5)$$

The velocity of the rocks with respect to Patroclus was the inverse of the velocity of Patroclus with respect to the rocks, namely, $v_R' = -u$.

Achilles approached him from his back, from the extreme of the other string with length *l'*. Achilles' velocity with respect to Patroclus can be calculated using the relativistic rule of addition of velocities and is given by:

$$v_A' = \frac{u}{1 - \frac{2u^2}{c^2}} \qquad (6)$$

The equations of movement of the rocks and Achilles in *S'* are:

$$x_R'(t') = \sqrt{1 - \frac{u^2}{c^2}}\, l - u t' \qquad (7.1)$$

$$x_A'(t') = -\sqrt{1 - \frac{u^2}{c^2}}\, l + \frac{u}{1 - \frac{2u^2}{c^2}} t' \qquad (7.2)$$

It can be seen in Eqs. (7.1) and (7.2) that the initial distances of Achilles and the rocks to Patroclus were the same. However, the speed of Achilles with respect to Patroclus was higher than the speed of the rocks. As a consequence, Patroclus observed how Achilles overtook him and encountered the rocks before he did. In this case, surprisingly, the arrival was not simultaneous.

To Achilles, the situation was somewhat similar. Initially he observed his friend at the end of the first string in front of him, whereas the rocks were twice as far, at the end of the second string. Again, the length of the strings (*l''*) contracted with respect to *l*:

$$l'' = \sqrt{1 - \frac{4u^2}{c^2}}\, l \qquad (8)$$

The rocks approached him with uniform velocity *v_R''=-2u*, while the velocity of Patroclus with respect to Achilles is given by:

$$v_P'' = -\frac{u}{1 - \frac{2u^2}{c^2}} \qquad (9)$$

The equations of movement of the rocks and Patroclus in *S''* are:

$$x_R''(t'') = 2\sqrt{1 - \frac{4u^2}{c^2}}\, l - 2ut'' \qquad (10.1)$$

$$x_P''(t'') = \sqrt{1 - \frac{4u^2}{c^2}}\, l - \frac{u}{1 - \frac{2u^2}{c^2}}\, t'' \qquad (10.2)$$

Eqs. (10.1) and (10.2) show that the distance covered by Patroclus in *S''* was half of the distance covered by the rocks. However, the speed of Patroclus in *S''* was higher than half the speed of the rocks. The encounters of Patroclus and the rocks with Achilles in *S''* were, again, not simultaneous.

For the sake of clarity, the equations of movement of Achilles, Patroclus and the rocks observed in *S*, *S'* and *S''* frames are depicted in Fig.3. It can be observed that the velocities of Achilles and the rocks measured in *S'* were not opposite, thus leading to the de-synchronization of the arrivals. In *S'* Achilles met Patroclus first, next Achilles met the rocks and finally the rocks met Patroclus. In *S''* the encounters follow the same temporal order as in *S'*. To Achilles, Patroclus arrived first to his position. Next, the rocks met him and finally Patroclus reached the rocks. The same is not true in *S*. To the observers on the rocks, the encounters were simultaneous.

**III. DISCUSSION OF THE RESULTS**

It has been shown how special relativity laws lead to the de-synchronization of the encounters with the rocks experienced by the two epic runners. To the observers on the rocks, the ratio of the distances between the runners and them was equal to the ratio of the runner's respective speeds and, therefore, their arrival was simultaneous. However, Patroclus and Achilles experienced that the ratio of the distances of the rocks and their contender to them was not equal to the ratio of their respective speeds. In essence, this feature arises from the limit imposed by the speed of light. No object in the universe can have a speed higher than *c* and this requirement has an effect on the rule that connects the relative velocities of an object in different inertial frames. It is worth noting that this rule is internally consistent. We have used the velocities of Patroclus and Achilles in *S* to calculate the velocities of the rocks and Achilles in *S'* as well as the velocities of the rocks and Patroclus in *S''*. Likewise, the velocities of the rocks and Patroclus in *S''* can be calculated from the velocities of the rocks and Achilles in *S'*. Additionally, the velocities of Patroclus and Achilles in *S* can be obtained from the corresponding velocities in *S'* or in *S''*.

The same cannot be said about the relativistic transformation of lengths measured by the different observers. We have calculated the lengths of the strings in *S'* and, therefore, the initial distances of the rocks and Achilles in *S'*, from the lengths of the strings determined in *S* (see eq.5). Next, we have calculated the lengths of the strings in *S''* also from the lengths of the strings in *S* (see eq.8). However, if we calculate the lengths of the strings in *S''* from the lengths of the strings previously determined in *S'* we obtain a different result, namely:

$$l'' = \sqrt{1 - \frac{\left(\frac{-u}{1-\frac{2u^2}{c^2}}\right)^2}{c^2}}\, l' = \sqrt{\left(1 - \frac{\frac{u^2}{c^2}}{\left(1-\frac{2u^2}{c^2}\right)^2}\right)\left(1 - \frac{u^2}{c^2}\right)}\, l \neq \sqrt{1 - \frac{4u^2}{c^2}}\, l \qquad (11)$$

It is also evident that the lengths of the strings measured in *S* (*l*) cannot be obtained from the lengths measured in *S'* (*l'*) or in *S''* (*l''*) using the formula of length contraction and the relative velocities of the corresponding inertial frames.

It is also instructive to analyze the time intervals measured by the different observers for the relative encounters of Achilles, Patroclus and the rocks. The time to encounter the rocks measured by Patroclus ($t_f'$) can be obtained from eq.(7.1). This time can be related to the time determined by the referees sat on the rocks for the same event ($t_f$) through the expression:

$$t_f = \frac{1}{\sqrt{1-\frac{v^2}{c^2}}}\, t_f' \qquad (12)$$

Similarly, according to Eqs. (10.1) and (1.3), the time span determined by Achilles when he encountered the rocks is related to $t_f$ through:

$$t_f = \frac{1}{\sqrt{1-\frac{4v^2}{c^2}}} t_f'' \qquad (13)$$

Eqs. (12) and (13) are the well known formulas showing the so called time dilation effect. To Patroclus and Achilles, the time elapsed before their encounter with the rocks was lower than that determined in *S*. Moreover, these time intervals can be related to the time interval measured in *S* using their relative velocities in the time dilation formulas. However, the validity of Eqs.(12) and (13) is somewhat accidental. It is derived from the reciprocal symmetry of the encounter of Patroclus and the rocks in *S* and *S'* as well as the encounter of Achilles and the rocks in *S* and *S''*. It is worth noting that a formula of time dilation like those expressed by Eqs. (12) and (13) cannot be used to relate the time intervals measured by the different observers for other events. For instance, the time interval determined by Achilles for his encounter with Patroclus cannot be related to the time determined by Patroclus for the same event using their relative velocity. Besides, it can be seen in Fig.3 that while the time measured by Achilles for his encounter with the rocks is lower than the time measured by Patroclus for the same event, the encounter of Patroclus and the rocks occurs later in *S''* than in *S'*.

**IV. CONCLUDING REMARKS**

Special relativistic effects have been exposed through a basic problem, represented in the framework of epic poetry. The selected problem shows that two events which are simultaneous in a given inertial frame lose that property in any other frame. This feature is characteristic of special relativity. In general, the temporal order of different events is not the same for every inertial observer. This means that time, in the sense we intuitively think of it as a superposition of ordered time intervals, is not universal. However, the notion of time under special relativity laws seems paradoxical to us. An alteration of the temporal order of events in the physical world opens a deep question on causality. Let us consider the case where, instead of running to the rocks, Achilles and Patroclus were running to catch a turtle resting on the rocks. Questions such as 'who caught the turtle?' would have different answers for different observers. But questions such as 'how must time be understood according to special relativity and what implications does this have on causality?', demand a great effort of attention and meditation to be solved.[2] To further analyze this kind of questions I propose a similar problem also inspired in the legend of Achilles and invite the interested reader solve the paradox.

*Achilles and Hector*

The legend tells that Achilles, enraged over the death of Patroclus, chased Hector, the son of Priam, around the wall of Troy. Hector was persuaded by Athena to stop running and he faced Achilles in a race. The runners should cover the distance of one stadium by running in opposite directions and reach a cross raised at the middle point of the line joining them (see Fig.4). A group of observers in rest relative to the sand and the cross watched the race. To them, the starts of Achilles and Hector and their arrivals to the cross were both simultaneous events. However, each one of the two runners perceived that he reached the cross before his opponent. So they felt cheated and decided to fight with weapons. The denouement of the story is well documented in the Iliad.[3]

## V. REFERENCES

[1] A particularly clear introduction to special relativity theory can be found in: Albert Einstein, "Relativity. The Special and the General Theory" (Penguin Classics, NY, 2006)

[2] See, for instance, the work by Kurt Gödel, "A remark about the relationship between relativity theory and idealistic philosophy", in *Albert Einstein, Philosopher-Scientist*, edited by Paul A. Schilpp (The Library of Living Philosophers, Evanston, Illinois, 1949, pp.555-562)

[3] http://classics.mit.edu/Homer/iliad.html

## VI. LIST OF SYMBOLS

$S$: inertial reference frame in rest relative to the strings and the rocks

$S'$: inertial reference frame in rest relative to Patroclus

$S''$: inertial reference frame in rest relative to Achilles

$c$: speed of light

$l$: length of the string in $S$.

$u$: speed of Patroclus in $S$

$x_P$: position of Patroclus in $S$

$x_A$: position of Achilles in $S$

$t$: coordinate time in $S$

$x_R'$: position of the rocks in $S'$

$x_A$': position of the Achilles in *S'*

*t'*: coordinate time in *S'*

$x_R$'': position of the rocks in *S''*

$x_P$'': position of Patroclus in *S''*

*t''*: coordinate time in *S''*

**CAPTIONS TO FIGURES**

**Fig.1.** Initial positions and velocities of: (a) Achilles and Patroclus, observed within the reference system *S*; (b) Achilles and the rocks, observed within the reference system *S'*; (c) Patroclus and the rocks observed within the reference system *S''*.

**Fig.2.** Graphical representation of the classical equations of movement of Patroclus (red), Achilles (blue) and the rocks (green) observed in *S* (a), *S'* (b) and *S''* (c) inertial frames. According to classical relativity, time can be defined by using a universal scale in the three inertial frames.

**Fig.3.** Graphical representation of the classical equations of movement of Patroclus (red), Achilles (blue) and the rocks (green) observed in *S* (a), *S'* (b) and *S''* (c) inertial frames. According to special relativity, each inertial frame has a different time scale, its coordinate time.

**Fig.4.** Initial positions and velocities of Achilles (to the left) and Hector (to the right), observed within the reference system in rest relative to the cross.

**FIGURES**

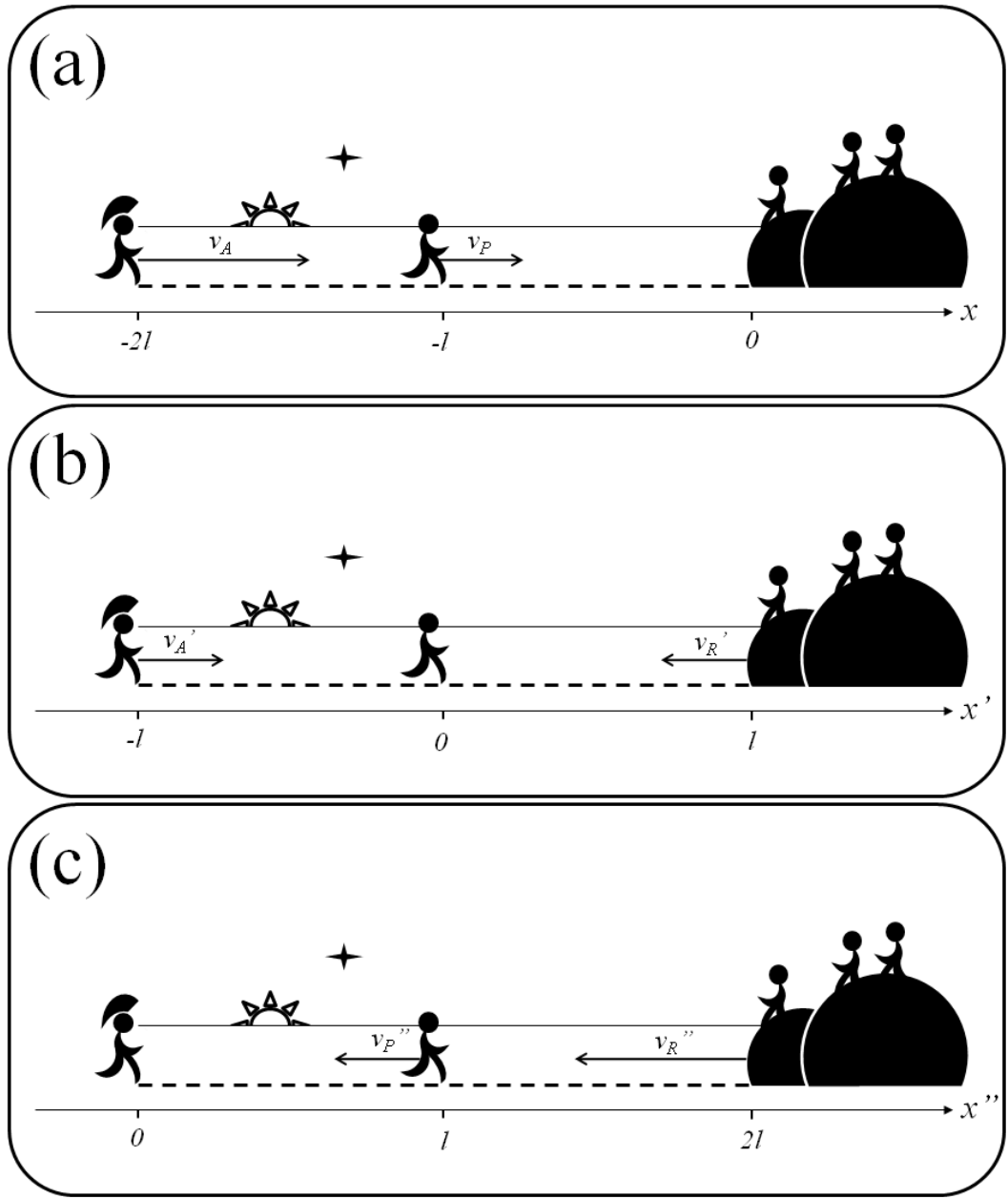

**Fig.1**

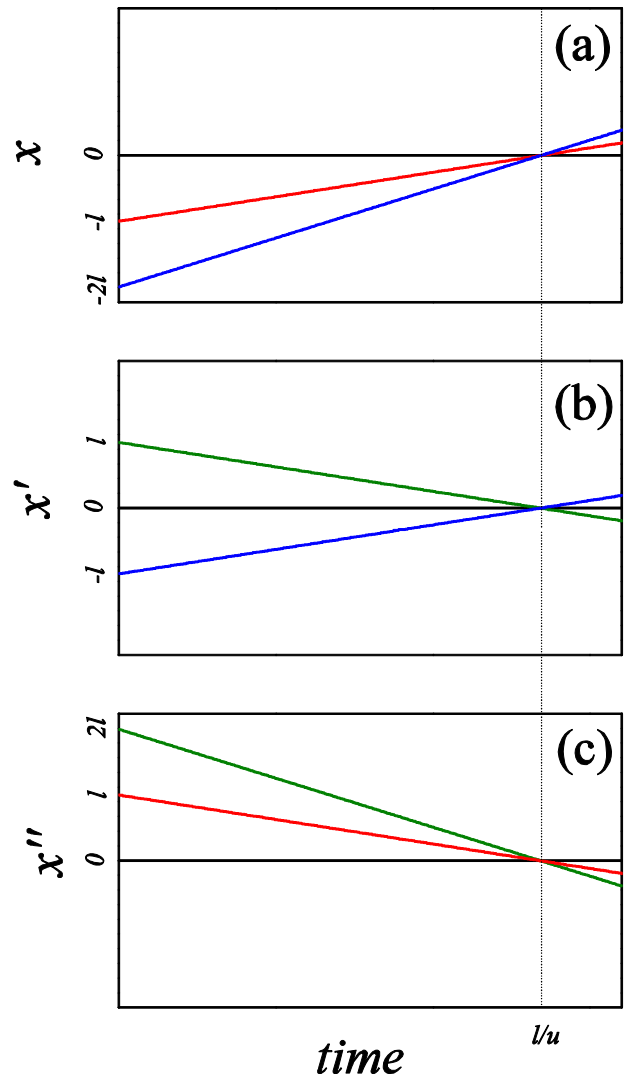

Fig.2

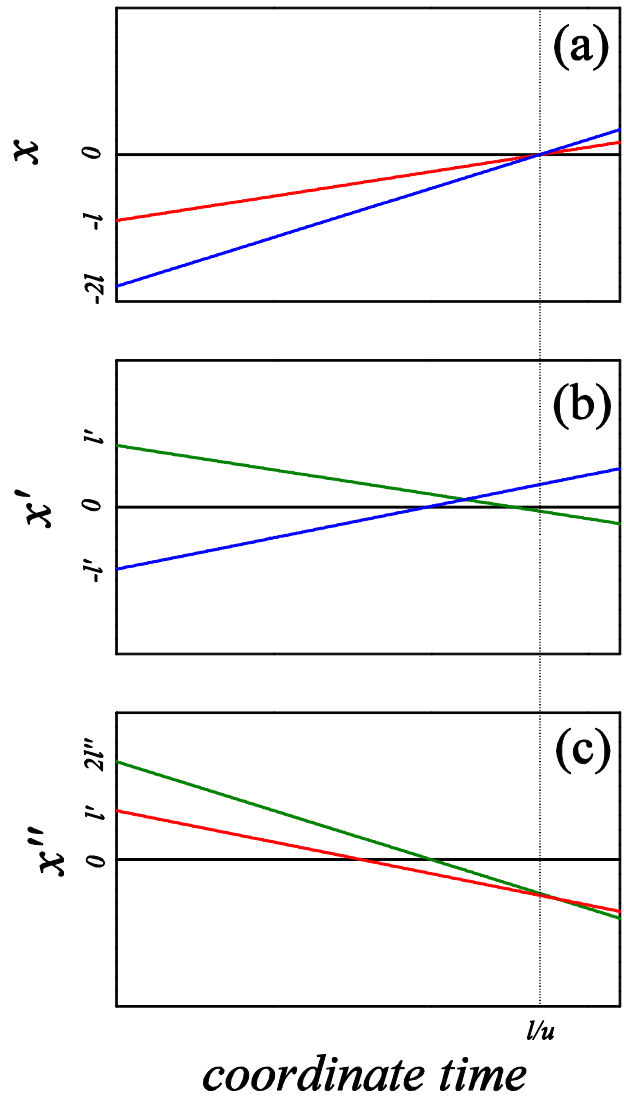

Fig.3

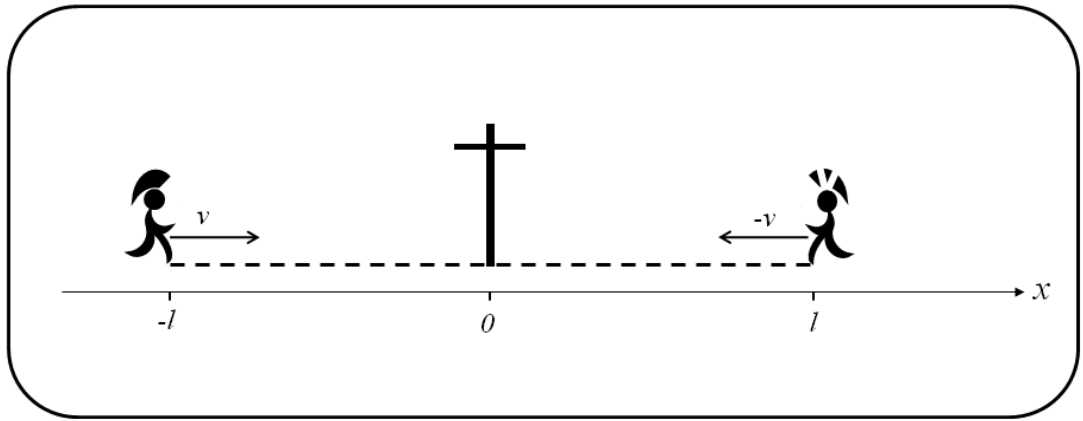

**Fig.4**